\documentstyle[epsf,psfig,prl,multicol,aps]{revtex}

\begin{document}
\title{Superconductivity Near Ferromagnetism in MgCNi$_3$} 
\author{H. Rosner,$^{1}$ R. Weht,$^{2,3}$ M. D. Johannes,$^{1}$ 
W. E. Pickett$^{1}$ and E. Tosatti$^{3,4}$ }
\address{
$^{1}$Department of Physics, University of California, Davis CA 95616 \\
$^{2}$Departamento de F\'{\i}sica, CNEA,
   Avda. General Paz y Constituyentes, 1650 - San Mart\'{\i}n, Argentina \\
$^{3}$ICTP, P.O. Box 586 34014 Trieste, Italy  \\
$^{4}$SISSA and INFM/SISSA, Via Beirut 2-4, 34014 Trieste, Italy}

\date{\today}
\maketitle
\begin{abstract}
An unusual quasi-two-dimensional heavy band mass van Hove singularity (vHs) 
lies very near the Fermi energy in MgCNi$_3$,
recently reported to superconduct at 8.5 K.  This compound is strongly
exchange enhanced and unstable to ferromagnetism upon hole doping with 
$\sim$12\% Mg$\rightarrow$Na or Li.  The 1/4-depleted fcc (frustrated)
Ni sublattice and lack of Fermi surface nesting argues against competing
antiferromagnetic and charge density wave instabilities.
We identify an essentially infinite mass along the M-$\Gamma$ line,
leading to quasi-two-dimensionality
of this vHs may promote unconventional $p$-wave pairing that could
coexist with superconductivity.  
\end{abstract}

\begin{multicols}{2}
The discovery of $\sim$40 K superconductivity in MgB$_2$~\cite{akimitsu} 
has spurred
interest in searching for superconductivity in unlikely materials, and
other discoveries are uncovering previously unanticipated relationships
between FM and superconducting states. 
Of this latter category, there
are now several examples, such as the magnetic organometallic 
(BETS)$_2$FeCl$_4$, where superconductivity is 
actually {\it induced}\cite{uji} by a strong
applied magnetic field rather than being destroyed by it, and 
the intermetallic UGe$_2$,
where superconductivity occurs\cite{saxena,huxley} in spite of 
strong ferromagnetism (FM) and
coexists with it to the lowest temperatures, and is almost certainly
triplet paired.  The question of   
FM\cite{sro1,sro2} on the surface of the exotic 
superconducting oxide Sr$_2$RuO$_4$ is yet another aspect of the strong
relationship between FM and superconductivity that is yet to be understood.

A new compound that would not
be considered a likely candidate for superconductivity
is MgCNi$_3$, whose conduction electrons
are primarily derived from Ni which is itself a FM, yet it
superconducts at 8.5 K.\cite{cava1}  This perovskite compound 
($a$ = 3.812 \AA) can be
regarded as fcc Ni that is expanded by 8\%, one quarter of the Ni replaced by 
Mg, then C atoms put into the octahedral sites.  Partial replacement of
Ni by Co or Cu reduces (or destroys)  
T$_c$,\cite{hayward} while other substitutions have not 
been reported.  The Hall coefficient is hole-like but strongly dependent
on temperature, unlike a conventional metal (Fermi liquid), while the
measured critical field H$_{c2}$ has a conventional shape.\cite{li}  
Tunneling spectra show a strong zero bias anomaly that has been
interpreted as evidence of strong coupling
superconductivity of an unconventional type.\cite{mao}

In this paper it is shown that this compound, in addition to being
superconducting, is also
an incipient FM, which can be driven to ferromagnetism by
partial ($\approx$12\%)
replacement of Mg with a monovalent metal such as Li or Na.  This close 
proximity of superconductivity to magnetism itself implies unconventional
pairing, and our identification of a van Hove singularity (vHs) 
with quasi-two-dimensional (2D) character
provides further support for unconventional pairing and possible coexistence of
FM and superconductivity, two types of collective order that are usually
antagonistic.

The perovskite structure itself
is rather unusual for such an intermetallic compound, 
since perovskites much more commonly have a strongly
negative ion (O$^{2-}$ or a negatively charged halide) on the site occupied
by Ni in this compound.  It is essential first to understand the character
of the charge carriers, for which purpose we have carried out full
potential, all-electron density functional based 
calculations.\cite{bandstr}
The resulting spectral distribution of the electronic states 
(for the experimental lattice constant $a$=3.812~\AA) is shown in
Fig.~1, and is indistinguishable from that presented by 
Singh and Mazin.\cite{singh}  
The conducting states at the Fermi level
are dominated by Ni $d_{xz}, d_{yz}$ and also $d_{x^2-y^2}$ character, in the 
local coordinate system in which the $\hat z$ axis is directed toward the two 
neighboring C atoms.  The remarkable feature of this compound is the 
sharp peak in the density of states (DOS) just 45 meV 
below the Fermi level.\cite{dugdale}  
This peak results from a van Hove singularity (vHs)
arising from a remarkably flat, primarily Ni $3d$ derived, band at and
around the $M = (1,1,0)\pi/a$ point in the simple cubic Brillouin zone.
Thus it is high mass Ni $3d$ holes
that form the superconducting pairs, and Ni-Ni hopping is important for
their transport.

To understand more clearly the origin of this peak, a fictitious material
$\Box ^{2+}$CNi$_3$ was studied, {\it i.e.} 
Mg was removed 
but its two valence electrons were retained.  While the C atom does have an
appreciable effect on the bonding and the resulting density of states (not
pictured), for practical purposes the Mg simply 
gives up its to valence electrons
to the bands (formed mainly by Ni) and has almost 
no other effect, as can be seen in Fig.~1.  We utilize 
this important point below.  If there were 0.5 electron less per cell,
E$_F$ would lie just at the peak in the DOS where N(E) is a
factor of eight larger. 

In addition to promoting superconductivity, a large value of N(E$_F$)
(equal to 2.4 states/eV-spin here)
leads an 
exchange-enhanced magnetic susceptibility $\chi$ that strongly opposes 
singlet superconductivity, or possibly even to a FM instability
(where $\chi \rightarrow \infty$) which is incompatible
with singlet superconductivity.  This latter scenario applies to 
Sr$_2$RuO$_4$,\cite{SRO}
which is a nearly FM superconductor (but only below 1.5 K) and is
now understood to be a parallel-spin-paired (triplet) superconductor.
Density functional
calculations are very reliable in calculating this tendency toward magnetism,
and indeed the instability to FM, particularly in intermetallic
compounds such as MgCNi$_3$.  The enhanced (observed) 
susceptibility\cite{janak} 
is given by
\begin{equation}
\chi = \frac{\chi_o}{1-N(E_F)I} \equiv S \chi_o,
\end{equation}
where $\chi_o$ = 2$\mu_B^2$N(E$_F$) is the bare susceptibility 
obtained directly from
the band structure 
and $I$ is the exchange interaction.  

We have calculated $I\approx 0.29\pm 0.01$ eV in two ways.  
One, which demonstrates 
directly our main thesis that MgCNi$_3$ is close to ferromagnetism,
was a calculation for ordered Mg$_{1/2}$Li$_{1/2}$CNi$_3$, which effectly
simply removes 0.5 valence electron from the cell.  This material is 
predicted to be
FM, and the exchange splitting $\Delta_{ex}$ between majority and minority
bands (Fig.~2) gives $I$= 0.30 eV from the relation 
$\Delta_{ex} = I m$ where $m$ is the 
FM moment in units of $\mu_B$.  
The other calculation of $I$ resulted
from fixed spin moment calculations,\cite{fixedspin} 
in which the energy $E(m)$ is calculated
subject to the moment being constrained to be $m$.  The behavior at small
$m$ is 
$E(m) = (1/2) \chi^{-1} m^2$
from which $I$ = 0.28 eV
can be extracted from Eq. (1).  
Singh and Mazin~\cite{singh} have obtained a similar value for the
Stoner parameter.
Thus $S$=3.3, and it is
certainly unexpected for a conventional (singlet) superconducting state 
to survive so near a FM instability, especially when the
superconducting carriers are heavy and are the same ones 
that will become magnetic.

To quantify how near this system is to being FM, we have carried out
(i) a series of virtual crystal calculations for 
Mg$_{1-\delta}$Na$_{\delta}$CNi$_3$ (justified by the results shown in
Fig. 1)
to find the concentration $\delta_{cr}$ of the FM critical point, 
and (ii) an extended Stoner
analysis.\cite{marcus}
The two results are consistent in predicting the onset of FM
at $\delta_{cr} \simeq 0.12$.  The 
ordered magnetic moment $m(\delta)$ versus hole doping level is shown  
in the inset in Fig.~3, where it is evident that, in the absence
of superconductivity, a moment
grows as $m(\delta) = {\cal G} 
      (\delta-\delta_{cr})^{1/2}$ for small $\delta-\delta_{cr}$
beyond the critical concentration.
The behavior of $m(\delta)$ in the small $m$ limit 
can be obtained analytically
from an expansion of the DOS $\bar N(m)$ averaged over the states within
$\pm m$ of $E_F$:
\begin{eqnarray}
\bar N(m,\delta) & \approx & \bar N(0,\delta_{cr})+
         \frac{d{\bar N}(0,\delta_{cr})}{d\delta}(\delta-\delta_{cr}) \\
& &    + \frac{1}{2} \frac{d^2{\bar N}(0,\delta_{cr})}{d^2m} m^2 = I^{-1}
\end{eqnarray}
and using $ \bar N(0,\delta_{cr}) = 1/I$ to obtain the square root law, with
\begin{equation}
{\cal G} = \Bigl| 2\frac{d{\bar N}(0,\delta_{cr})}{d\delta} /
    \frac{d^2{\bar N}(0,\delta_{cr})}{d^2m} \Bigr|^{1/2}\approx 1.7 \mu_B.
\end{equation}

There are experimental indications from tunneling\cite{mao} 
that the superconductivity in
MgCNi$_3$ may arise from triplet pairing.\cite{tunnel}  
Our results show that hole doping
with Na or Li will be an excellent way to probe this possibility.  While it
is unexpected that singlet superconductivity 
would occur at all in a Ni compound
that is as strongly exchange enhanced as MgCNi$_3$ is 
(the increasing enhancement that diverges as 
$\delta \rightarrow \delta_{cr}$ should
kill singlet superconductivity very quickly),
for triplet pairing
the increasing magnetic correlations may provide the coupling (as recently
argued for heavy fermion superconductors\cite{HFsc}) and enhance T$_c$
as $\delta \rightarrow \delta_{cr}$.  

We now consider in more detail the electronic structure and its implications,
especially in the critical region $\delta \approx 0.12$.  In this region
E$_F$ is still 35 meV from the vHs, so (near) divergence of N($\varepsilon$)
{\it per se} is not a crucial consideration.
However, FM, AFM, and pairing susceptibilities should be considered,
and they
will involve energy denominators
$\delta \varepsilon = \varepsilon_{\vec k,n} - 
\varepsilon_{\vec k+\vec q,m}$, where $n, m$ label the
three M points where the vHs occur.  For $n=m$, $\delta \varepsilon$ 
will be small for small $\vec q$,
which relates to the FM instability we have considered above.
Since there
are vHs at three inequivalent M points, there will also be a
potential AFM instability
near a vector $\vec Q$ that spans two inequivalent vHs.  These
values of $\vec Q$ are in fact equal to M, so AFM tendencies are peaked 
at $\vec Q = (1,1,0)\pi/a$ and symmetrically related points.  

The dominant instability is determined largely by phase space availability.
The band giving the vHs is quite flat (to within 50 meV)
in a roughly cubic region of
side $\pi/a$ centered on each M point, which totals to 3/8 of the Brillouin
zone volume.  Band masses at the vHs are quite large: $m^*\approx$~15 along 
$M-X$, $m^*\approx$ -15
along $M-R$.  Along $M-\Gamma$, however, 
this band is constant to within 0.2 meV for a distance of $\approx \frac
{1}{3}\frac{\pi}{a}$,
{\it i.e.} it has effectively
an infinite effective mass along this line.  The dispersion relation in the
$X-M-\Gamma$ plane is pictured in Fig.~4.  Since the plane containing this
dispersionless line is oriented differently for each M point, 
there is no pronounced
nesting of the vHs.  However, because of the lack of dispersion along one
direction,
the shape of the N(E)
peak in Fig. 1 has the logarithmic divergence characteristic of 2D models.  
This vHs is remarkably {\it insensitive} to volume, being essentially
unchanged relative to E$_F$ when the lattice constant is decreased by 5\%, 
which makes pressure a relatively ineffective tool (except perhaps to tune the
Curie temperature beyond the critical point).

The structure of the vHs suggests that the resulting physics could be a 
three dimensional
generalization of the 2D vHs models.
The strong exchange enhancement and the proximity to a FM instability
is most consistent with triplet pairing, appropriate forms of
which can coexist with FM as in UGe$_2$, or with incipient FM as in
Sr$_2$RuO$_4$.
There are numerous possibilities of $p$ (or higher odd)
symmetry for a cubic system\cite{rudd}, some of which have gaps but
many of which have nodes and therefore
are gapless.  It typically requires high quality thermodynamic and 
spectroscopic data, which are not yet available, to determine the 
character of the order parameter.

The theoretical case for superconductivity arising
from longitudinal magnetic fluctuations was 
laid out by Fay and Appel,\cite{Fay} who
found that T$_c$ should peak near $\delta=\delta_{cr}$ (on either side)
but would vanish at the critical point.  
Machida and Ohmi\cite{machida}
emphasize that a non-unitary triplet state is most likely in a case
that may apply to MgCNi$_3$, since it 
intrinsically breaks time-reversal symmetry as
does ferromagnetism.  In such non-unitary phases, a magnetic field may
enhance T$_c$.\cite{machida}

The most
intriguing possibility is that (triplet) superconductivity might coexist
with ferromagnetism, as reported recently for 
UGe$_2$.\cite{saxena,huxley,shick}  
Mg$_{1-\delta}$(Na,Li)$_{\delta}$CNi$_3$ provides the
new possibility, if indeed coexistence occurs due to triplet pairing,
of studying the emergence of itinerant ferromagnetism (as $\delta$ crosses
$\delta_{cr}$) within an existing
superconducting phase.  The phenomenological theory of coexistence in
just such a case has been put forward recently,\cite{karchev} concluding
that the heat capacity has a linear-in-T term that is strongly dependent
on the magnetization.  Solutions of the Eliashberg equations for a spin
fluctuation system near the quantum critical point\cite{bedell} 
suggest that triplet
superconductivity might not be as strongly favored near the critical point
as might have been anticipated.
Hole-doped MgCNi$_3$ appears to be an excellent system to use as a probe
of these fundamental questions.

The FM instability at $\delta=\delta_{cr}$ was demonstrated above.
Can AFM ordering of the Ni atoms, such as is encouraged by inequivalent vHs
in other systems, occur here?  The Ni sublattice is a 1/4
depleted fcc lattice, but the
periodic depletion does not alter the well known frustration of N\'eel
order on the fcc lattice.  Thus while short range AFM 
correlations may be strong,
AFM {\it ordering} will be opposed by frustration.

To summarize, we have shown that the superconductor MgCNi$_3$ is near
a ferromagnetic instability that can be reached by hole doping on the Mg site.
The effective carriers are Ni-derived holes of very high band mass (likely
enhanced by dynamic spin fluctuations and phonons).
The FM instability is related to an unusual quasi-2D heavy
mass van Hove singularity less that 50 meV below E$_F$.  This quasi-2D
character supports earlier suggestions that its superconductivity is
unconventional in nature, and we suggest that hole-doping is an ideal
way to probe the onset of ferromagnetism in the superconducting state. 

W.E.P. and E.T. are grateful to R. Cava for conversations and for
early communication of research results.
This work was supported by the Italian MURST COFIN, by 
European Union TMR Contract 
ERBFMRXCT970155, by U. S. Office of Naval Research 
Grant N00017-97-1-0956, by National Science Foundation Grant DMR-9802076,
and by the Deutscher Akademischer Austauschdienst.
RW aknowkedges support from Fundaci\'on Antorchas Grant No 13661-27.

\vskip -5mm

\begin{figure}[tbp]
\epsfxsize=7.0cm\centerline{\epsffile{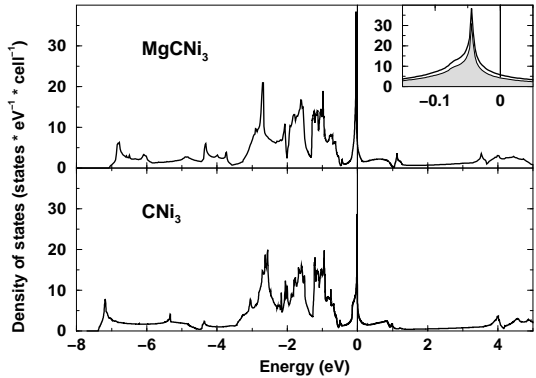}}
\caption{
The density of states of MgCNi$_3$ (top panel), showing the
nearly filled Ni $d$ states and the extremely sharp and
narrow peak just below E$_F$
arising from a van Hove singularity.
The inset gives a blowup of the peak, with the shaded portion
indicating the dominant Ni $3d$ contribution.
The lower panel shows the density of states for $\Box ^{2+}$CNi$_3$
(see text).
}
\label{Bands1}
\end{figure}

\begin{figure}
\epsfxsize=7.0cm\centerline{\epsffile{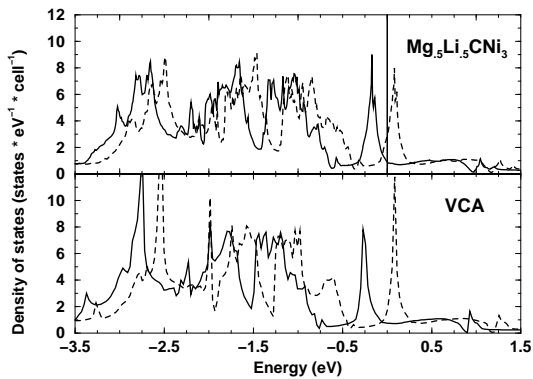}}
\caption{{\bf Ferromagnetic states obtained from self-consistent
calculation.}
Density of states of ferromagnetic $\delta=0.5$ materials. 
Top panel:
the ordered compound Mg$_{0.5}$Li$_{0.5}$CNi$_3$.  Bottom panel: a virtual
crystal result for Mg$_{0.5}$Na$_{0.5}$CNi$_3$. This level of
doping results in a filled majority van Hove peak and an empty minority 
van Hove peak.  
The magnetic moments are
0.83 $\mu_B$ and 0.95 $\mu_B$, respectively.
}
\end{figure}

\begin{figure}[t]
\epsfxsize=7.0cm\centerline{\epsffile{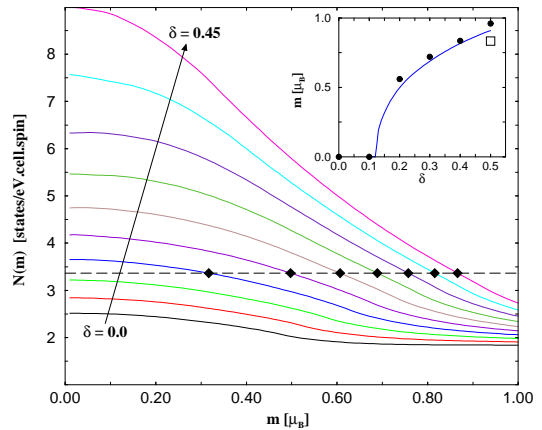}}
\caption{ 
The mean value N(m) of the density of states around $E_F$
necessary to create a magnetic moment $m$, versus $m$.  Hole-doping
concentrations $0\leq \delta \leq 0.45$ are shown.  The dashed line indicates
1/$I$ ($I$=0.29 eV). 
Solid curves give
results from a rigid band treatment based on the MgCNi$_3$ ($\delta=0$)
DOS.
The inset gives the predicted value of the ferromagnetic moment versus
the hole concentration from the Stoner model (solid line) and from
specific self-consistent virtual crystal calculations (dots)
for Mg$_{1-\delta}$Na$_{\delta}$CNi$_3$, which indicates the
consistency.  The square gives the moment for the ordered compound
Mg$_{1/2}$Li$_{1/2}$CNi$_3$ discussed in the text and in Fig.~2.
}
\label{DOS}
\end{figure}

\begin{figure}[t]
\epsfxsize=7.0cm\centerline{\epsffile{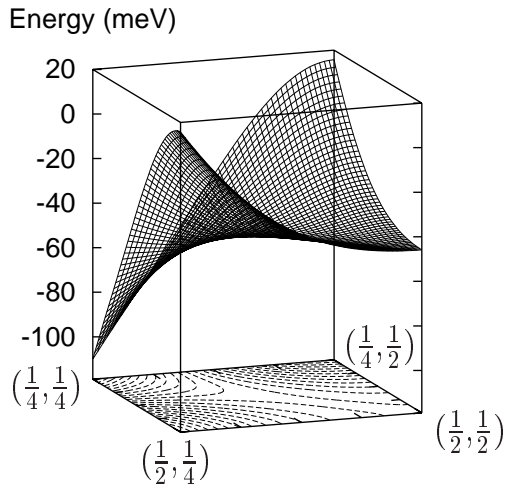}}
\caption{ 
Surface plot (and contour plot below) of the van Hove 
singularity in $\varepsilon_k$ (relative to E$_F$)
in the $\Gamma-M-X$ plane,
with $M$ at the right hand corner (planar coordinates are given in units
of $2\pi/a$).  Note the extreme flatness of the
dispersion along the diagonal $M\rightarrow\Gamma$.  
The negative effective mass sheet of
the vHs lies in the third direction.
}
\label{vHs}
\end{figure}
\end{multicols}

\begin{references}
\bibitem{akimitsu}J. Nagamatsu {\it et al.}, Nature {\bf 410}, 63 (2001).
\bibitem{uji}S. Uji {\it et al.}, Nature {\bf 410}, 908 (2001).
\bibitem{saxena} S.S. Saxena {\it et al.},
          {\it Nature} {\bf 406}, 587  (2000). 
\bibitem{huxley}A. Huxley {\it et al.}
          Phys. Rev. B {\bf 63}, 144519 (2001).
\bibitem{sro1}R. Matzdorf {\it et al.}, Science {\bf 289}, 746 (2000).
\bibitem{sro2}K. M. Shen {\it et al.}, cond-mat/0105487.
\bibitem{cava1}T. He {\it et al.}, Nature {\bf 411}, 54 (2001).
\bibitem{hayward}M. A. Hayward {\it et al.}, cond-mat/0104541.

\bibitem{li}S. Y. Li {\it et al.}, cond-mat/0104554.

\bibitem{mao}Z. Q. Mao {\it et al.}, cond-mat/0105280.
%
\bibitem{bandstr}We used two well established methods of calculation:
  WIEN97, see P. Blaha, K. Schwarz, and J. Luitz, Vienna
  University of Technology, 1997, improved and updated version of the
  original copyrighted WIEN code, which was published by P. Blaha,
  K. Schwarz, P. Sorantin, and S. B. Trickey, Comput. Phys. Commun.
  {\bf 59}, 399 (1990);
  a full potential nonorthogonal minimum basis
  local orbital method described in K. Koepernik and H. Eschrig,
  Phys. Rev. B {\bf 59}, 1743 (1999).  
\bibitem{singh}D. J. Singh and I. I. Mazin, cond-mat/0105577.
\bibitem{dugdale}Our band structure results differ somewhat
 in the sharpness and position of the vHs from other reports
(A. Szajek, J. Phys.: Condens. Matt. {\bf 13}, L595 (2001);
 S. B. Dugdale and T. Jarlborg, cond-mat/0105349;
 J. H. Shim, B. I. Min, cond-mat/0105492)
 probably because we have used full potential methods, and more than 800
 points in the irreducible zone were used to calculate the DOS.
\bibitem{SRO}Y. Maeno {\it et al.}, Nature {\bf 372}, 532 (1994).
 I. I. Mazin and D. J. Singh, Phys. Rev. Lett. {\bf 79}, 733 (1997).
\bibitem{janak} J. F. Janak, Phys. Rev. B {\bf16}, 255 (1977).
\bibitem{fixedspin}K. Schwarz and P. Mohn, J. Phys. F {\bf 14}, L129 (1984).
\bibitem{marcus}P. M. Marcus and V. L. Moruzzi, Phys. Rev. B {\bf 38},
   6949 (1988).
\bibitem{tunnel}Since doping simply due to the presence of an interface
may occur at a tunnel junction, and its requires only a small amount of
hole doping to introduce Ni moments, the possibility of magnetic moments at
the interface should not be overlooked.
\bibitem{HFsc}N. D. Mathur {\it et al.}, Nature {\bf 394}, 39 (1998).
\bibitem{rudd}R. E. Rudd and W. E. Pickett, Phys. Rev. B {\bf 57}, 557 (1998),
 and references therein.
\bibitem{shick}A. B. Shick and W. E. Pickett, Phys. Rev. Lett. 
   {\bf 86}, 300 (2001).
\bibitem{karchev}N. I. Karchev {\it et al.}, Phys. Rev. Lett. {\bf 86},
    846 (2001).
\bibitem{bedell}Z. Wang {\it et al.}, cond-mat/0104097.
\bibitem{Fay} D. Fay and J. Appel, Phys. Rev. B {\bf 22}, 3173 (1980).
\bibitem{machida}K. Machida and T. Ohmi, Phys. Rev. Lett. {\bf 86},
   850 (2001).
\end{references}
\end{document}